\documentclass[reprint, prb, superscriptaddress, float, floatfix, longbibliography,citeautoscript]{revtex4-1}
\usepackage{times}
\usepackage{sidecap}
\usepackage{amsmath}
\usepackage{amssymb}
\usepackage{dsfont}
\usepackage{graphicx}
\usepackage{epstopdf}
\usepackage[hidelinks]{hyperref}
\hypersetup{colorlinks = true, linkcolor = blue, anchorcolor = blue, citecolor = blue, pdfstartpage = 1, filecolor = red, urlcolor = black}
\usepackage{feynmf}
\usepackage{chngcntr}

\usepackage{xcolor}
\usepackage{xspace}
\usepackage{marginnote}

\newcommand{\units}[1]{{\,\rm #1}}

\newcommand{\fig}[1]{Fig.\,\ref{#1}} 
\newcommand{\Fig}[1]{Figure\,\ref{#1}}

\newcommand{\eq}[1]{Eq.\,(\ref{#1})}

\newcommand{\eqsand}[2]{Eqs.\,(\ref{#1}) and (\ref{#2})}

\newcommand{\be}{\begin{equation}}
\newcommand{\ee}{\end{equation}}
\newcommand{\bea}{\begin{eqnarray}}
\newcommand{\eea}{\end{eqnarray}}

\newcommand{\lb}{\left[}
\newcommand{\rb}{\right]}

\newcommand{\la}{\langle}
\newcommand{\ra}{\rangle}

\def\Re{\operatorname{Re}}
\def\Im{\operatorname{Im}}

\def\r{\mathbf{r}}
\def\k{\mathbf{k}}
\def\q{\mathbf{q}}
\def\d{\mathrm{d}}
\def\w{\omega}
\newcommand{\dOm}{\delta\Omega}
\newcommand{\Rinf}{R_2^\infty}
\newcommand{\Rm}{R_2^{\rm mol}}
\newcommand{\Dinf}{D_\infty}
\newcommand{\G}{\mathcal{G}}

\begin{document}

\title{Observation of magnetic structural universality using transverse NMR relaxation}
%\title{Transverse NMR relaxation characterizes mesoscopic structural organization}

\author{Alexander Ruh}
\affiliation{Medical Physics, Department of Radiology, Medical Center - University of Freiburg, Faculty of Medicine, University of Freiburg, Freiburg, Germany}
\affiliation{Department of Radiology, Feinberg School of Medicine, Northwestern University, Chicago, IL, USA}

\author{Philipp Emerich}
\affiliation{Medical Physics, Department of Radiology, Medical Center - University of Freiburg, Faculty of Medicine, University of Freiburg, Freiburg, Germany}

\author{Harald Scherer}
\affiliation{Institute of Inorganic and Analytical Chemistry, University of Freiburg, Freiburg, Germany}

\author{Dmitry S. Novikov}
\affiliation{Bernard and Irene Schwartz Center for Biomedical Imaging, Department of Radiology, New York University School of Medicine, New York, NY, USA}

\author{Valerij G. Kiselev}
\email{valerij.kiselev@uniklinik-freiburg.de}
\affiliation{Medical Physics, Department of Radiology, Medical Center - University of Freiburg, Faculty of Medicine, University of Freiburg, Freiburg, Germany}

\date{\today}

\begin{abstract}
\noindent
Transverse NMR relaxation from spins diffusing through a random magnetic medium is sensitive to its structure on a mesoscopic scale. In particular, this results in the time-dependent relaxation rate. We show analytically and numerically that this rate approaches the long-time limit in a power-law fashion, with the exponent reflecting the disorder class of mesoscopic magnetic structure. The spectral line shape acquires a corresponding non-analytic power law singularity at zero frequency. 
We experimentally detect a change in the dynamical exponent as a result of the transition into a maximally random jammed state characterized by hyperuniform correlations. 
\end{abstract}

\maketitle

{%\bf
\noindent
Transverse NMR or EPR relaxation is sensitive to the spectral density of the fluctuating environment.\cite{bloembergen48,anderson53,Khaetskii2002,Uhrig2007}
Such environment can emerge when the spins 
travel across magnetically disordered media, 
such as semiconductors,\cite{Dyakonov71,Salis2001} porous rocks,\cite{hurlimann98,song2000} and biological tissues.\cite{glasel74,thulborn82,gillis87,Weisskoff94,Davis2018,kiselev2018_review}
While the magnetic structure remains static, the stochastic motion of spins enables sampling its spatial fluctuations, resulting in a non-Lorentzian spectral lineshape and a time-dependent relaxation rate.\cite{yablonskiy94,kiselev98,jensen2000_dnr,kiselev2002,sukstanskii2003,sukstanskii2004,novikov2008}

Here we show that this rate approaches the long-time limit in a power-law fashion, and relate its 
dynamical exponent to the structural exponent\cite{novikov2014} characterizing long-range spatial correlations of magnetic structure. 
In particular, we experimentally observe a change in the dynamical exponent as a result of the transition into a maximally random jammed state \cite{torquato2000} characterized by hyperuniform correlations.\cite{torquato2003,donev2005_PRL}
Our results reflect the hierarchical nature of structural complexity contributing to a macroscopic NMR signal: its functional form is defined by the structural universality class, whereas microscopic parameters affect the nonuniversal coefficients. The relation between relaxational dynamics and magnetic structure opens the way for noninvasive characterization of porous media, complex materials and biological tissues. 
}

The transverse NMR relaxation signal is given by the average of the precession phase factor 
$s(t) = \langle e^{-i \varphi(t)} \rangle$, $\varphi = \int_0^t \Omega(\tau)\, \d\tau$, where $\Omega(\tau)$ is the fluctuating Larmor frequency offset experienced by nuclear spins. This averaging for times $t\gg t_c$ exceeding the correlation time $t_c$ of $\Omega(t)$ falls into the realm of the central limit theorem:\cite{anderson53,Dyakonov71} the signal is asymptotically determined by its second-order cumulant 
$\ln s(t) \approx -\langle \varphi^2 \rangle / 2 \sim - \langle \varphi_1^2\rangle \cdot (t/t_c)$, where 
$\langle \varphi_1^2\rangle$ is the phase variance on a single {correlated} ``step", and these variances add up on the path split into a large number $\sim t/t_c$ of uncorrelated steps. For a weak dephasing, 
the resulting relaxation rate $R_2\sim \langle \varphi_1^2\rangle/t_c \sim \langle \Omega^2\rangle t_c$ is the essence of the motional-narrowing picture of Anderson and Weiss,\cite{anderson53} which applies to the dipole-dipole interaction between the excited spins on a molecular scale. Such monoexponential molecular relaxation emerges from the enormous separation of scales between the correlation time $t_c\sim 1-10\,$ps of molecular motion and the typical NMR measurement time scale $t\sim 1-100\,$ms. 

In this work, we consider the loss of Larmor precession coherence in media with a static magnetic structure on a much larger, \textit{mesoscopic} scale \cite{glasel74,thulborn82,gillis87} relevant for NMR experiments in porous rocks \cite{hurlimann98,song2000} and in biological tissues.\cite{Weisskoff94,yablonskiy94,kiselev98,jensen2000_dnr,kiselev2002,sukstanskii2003,sukstanskii2004,novikov2008,Davis2018,kiselev2018_review}
Here, the individual precession phases 
$\varphi(t) = \int_0^t \Omega(\r_\tau) \, \d\tau$ decohere due to the path-dependent Larmor frequency offset $\Omega(\r_\tau)$ on their Brownian trajectories $\r_\tau$ induced by heterogeneous medium's magnetic susceptibility. The macroscopic $t\to\infty$ rate $\Rinf \sim \langle \Omega^2(\r)\rangle\, t_c$ decreases for a faster diffusion constant $D$, as the time $t_c \sim l_c^2/D$ to travel across the disorder correlation length $l_c$ shortens, exemplifying the diffusion narrowing.\cite{kennan94,kiselev98,jensen2000_dnr,kiselev2018_review} 
Importantly, the typical mesoscopic correlation times $t_c$ can be of the order of the NMR measurement time scale, which makes it possible to explore the {\it transient} signal evolution $s(t)$ before the long-time monoexponential limit is reached, via studying the corresponding time-dependent relaxation rate $R_2(t) = -\d\ln s(t)/\d t$.

%%%%%%%%%%%%%%%%%%%%%%%
\begin{figure*}
\centering
\includegraphics[width=\linewidth]{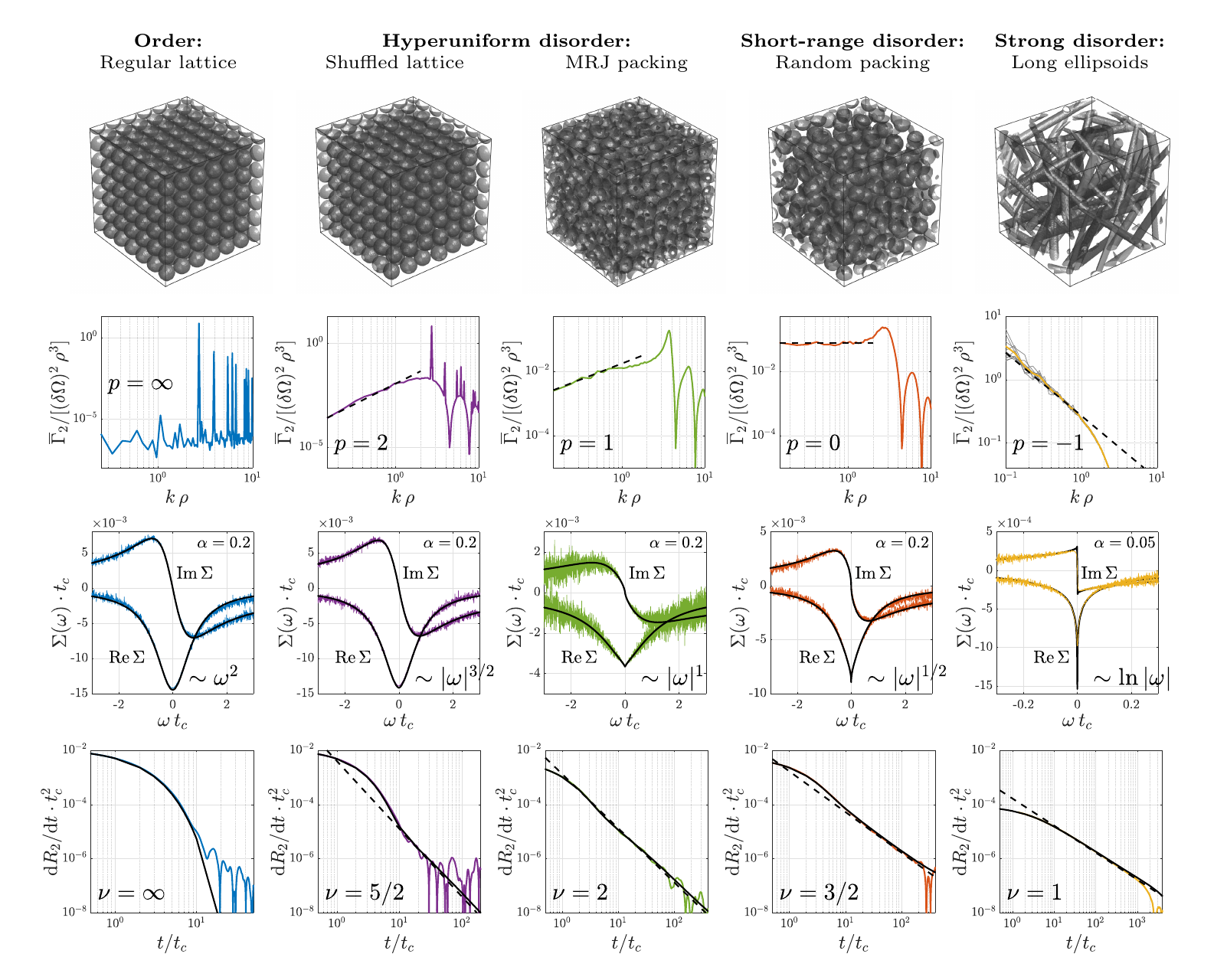}
\vspace{-10mm}
\caption{{\bf Illustration of the universality relations (\ref{dR2}) -- (\ref{Gamma_chi}) for the disorder classes in $d=3$ dimensions.}
Columns show the results for five fully permeable synthetic media with qualitatively distinct structural fluctuations 
(see Methods). 
First row shows a quarter of the Monte Carlo simulation box in each dimension; second row shows the angular-averaged power spectra $\overline \Gamma_2(k)$, \eq{Gamma_k}; 
third and fourth rows show the self-energy part $\Sigma(\omega)$ (see text) and $\d R_2(t) / \d t$, 
numerically obtained from the simulations. 
The packing of long ellipsoids has large fluctuations at small $k$. 
This is a finite-size effect that is alleviated by the ensemble averaging (thicker yellow line) over 10 disorder realizations (thin gray lines). 
$\Sigma(\omega)$ and $\d R_2(t) / \d t$ obtained from MC simulations (colors) are compared with the leading-order calculation (black lines) by integrating the numerically found power spectra according to \eqsand{sigCorrection}{dR2pert}, respectively, while neglecting $\Rinf$ on the right-hand side. For $\d R_2(t) / \d t$ (fourth row) this agrees well with the asymptotic limit (\ref{dR2}) with the exponent (\ref{nu}) (dashed lines). Parameters $\rho$, $\dOm$, $\alpha$, and $t_c$ are defined in Methods.}
\label{fcorrdR2}
\end{figure*}
%%%%%%%%%%%%%%%%%%%%%%%

The mesoscopic contribution to the transverse relaxation, in principle, depends on myriads of parameters characterizing the spatial organization of susceptibility-induced $\Omega(\r)$. It is generally non-universal, and sensitive to the shape of magnetic inclusions (e.g.\ cells).\cite{gillis95,kiselev2002} Our main result is the {\it universal} feature of the mesoscopic relaxation which manifests itself in the power-law tail in the approach of $R_2(t)$ to $\Rinf$,
\begin{equation} \label{dR2}
\frac{\d R_2 }{\d t} \sim t^{-\nu} \quad \text{for} \quad t\gg t_c
\end{equation}
such that $ \Rinf - R_2(t) \sim t^{-\nu+1}$ for $\nu>1$, and $R_2(t) \sim \ln t$ for $\nu=1$; an upper temporal limit on this behavior is discussed below. We relate the dynamical relaxation exponent 
\begin{equation} 
\nu = {p+d \over 2}
\label{nu} 
\end{equation}
to the large-scale statistics of the structural organization: The relevant signature of the $d$-dimensional medium, represented by magnetic susceptibility $\chi(\r)$ varying on the mesoscopic scale, is embodied by the (magnetic) structural exponent $p$, which we define via the $k\to0$ scaling of the power spectrum
\be \label{Gamma_chi}
\Gamma_2^\chi(k) =\int\!\d^d\r\, e^{-i\k\r}\, \langle \chi(\r_0+\r)\chi(\r_0)\rangle_{\r_0} 
\sim k^p \,, \quad k\to 0 \,.
\ee
The exponent $p$ takes a few discrete values,\cite{novikov2014} characterizing distinct universality classes of structural disorder. 

\begin{figure}[b]
\includegraphics[width=\linewidth]{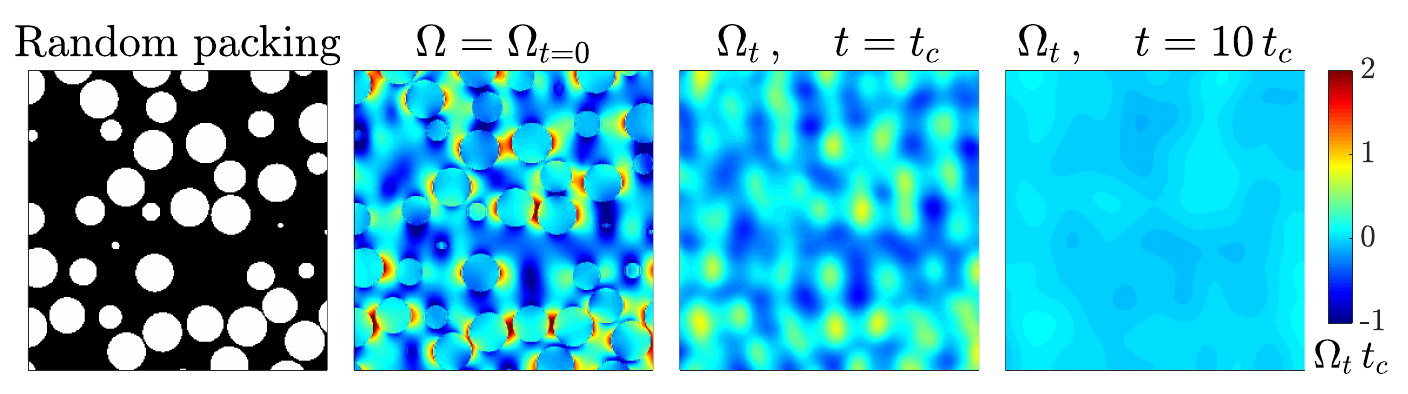}
\vspace{-6mm}
\caption{{\bf Larmor frequency offset and its coarse-graining by diffusion.} A section through the 3d random sphere packing from the fourth column in \fig{fcorrdR2}: The Larmor frequency $\Omega(\r) \equiv \Omega_{t=0}(\r)$ induced by the distinct magnetic susceptibility assigned to the spheres, and the coarse-grained $\Omega_t (\r)$ for $t=t_c$ and $10\, t_c$. The coarse-graining occurs via the local averaging of the frequency over the length $L(t)\sim \sqrt{Dt}$.}
\label{fieldfilter} 
\end{figure}

The key relation (\ref{nu}) is illustrated in \fig{fcorrdR2} for five statistically isotropic disorder classes in $d=3$ dimensions using Monte Carlo (MC) simulations, where we identify the exponent $p$ in the angular-averaged power spectra 
\begin{equation}\label{Gamma_k} 
\overline\Gamma_2(k) = \frac{\langle \Omega_{-\k} \Omega_\k \rangle_\mathbf{\hat k}}{V} 
= c_d (4\pi \gamma B_0)^2 \cdot \Gamma_2^\chi(k) \sim k^p
\end{equation}
of the susceptibility-induced Larmor frequency offset\cite{deville79,kiselev2002}
$\Omega_\k = 4\pi \gamma B_0 \chi_\k Y_{\hat\k} $ (\fig{fieldfilter}), where $Y_{\hat\k} = 1/3 - k_z^2/k^2$ is the elementary dipole field, $\gamma B_0$ is the average Larmor frequency, and $V$ is the sample volume. The scaling of the power spectra of the structure and of the induced frequency with $k=|\k|$ is similar\cite{novikov2008} due to the $Y(\r) \sim 1/r^d$ dependence of the dipole field, with $c_3 = \langle |Y_{\hat\k}|^2\rangle_{\hat\k} = {4/45}$, meaning that the transverse relaxation effectively samples the structure of the medium (i.e.\ $\chi(\r)$) {\it directly}, even though it senses the induced $\Omega(\r)$.

%%%%%%%%%%%%%%%%%%%%%%%
%\begin{SCfigure*}[0.5][h!!]
%\end{SCfigure*}

\begin{figure*}[t]
\includegraphics[width=\textwidth]{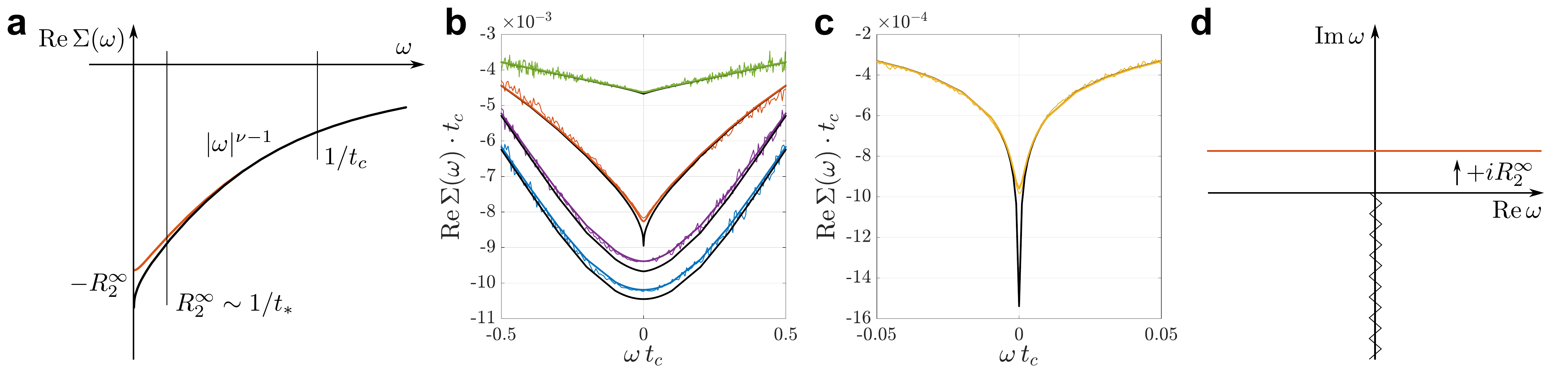}
\vspace{-7mm}
\caption{{\bf Self-consistent regularization of the $\omega\to0$ behavior.} Panel (a) shows a schematic of $\Sigma(\w)$ for small $\w$. Panels (b,c) show the zoomed low-frequency range from \fig{fcorrdR2} using the same colors (lines vertically shifted for the readability). The thick noiseless colored curves depict the self-consistent correction, \eq{sigCorrection}, to the leading-order calculation (black lines), resulting in a perfect agreement with simulations. Panel (d) depicts the sampling of $\Sigma(\w)$ off the singularity at $\w=0$ due to the self-regularization for long times that results in a finite $\Rinf$.}
\label{fig:selfconsist} 
\end{figure*}
%%%%%%%%%%%%%%%%%%%%%%%

The universality (\ref{nu}) in the diffusion-narrowing regime can be used as a probe of the global structural organization of magnetically heterogeneous media, and for the mesoscopic model selection. 
In \fig{fcorrdR2}, four kinds of identical sphere arrangements, and one with randomly placed long (prolate) ellipsoids, represent five distinct universality classes. 
In particular: 
\textit{Order}, represented by a cubic lattice of spheres, shows no long-range fluctuations and $\overline\Gamma_2(k) \equiv 0$ for small $k$, which can be associated with an exponent $p=\infty$, yielding an exponentially fast decay of $\d R_2/\d t$ (faster than any inverse power law). 
For \textit{hyperuniform disorder},\cite{torquato2003} $p>0$, these fluctuations are not completely absent but are suppressed. 
Examples are a \textit{shuffled lattice},\cite{gabrielli2002} where the lattice objects are randomly displaced from their original positions, showing a quadratic behavior $\overline\Gamma_2|_{k\to0} \sim k^2$, with $p=2$ yielding $\nu=5/2$; and a 
\textit{maximally random jammed (MRJ) packing},\cite{torquato2000} where $\overline\Gamma_2|_{k\to0} \sim k$, with a nontrivial exponent\cite{donev2005_PRL} $p=1$, which manifests itself in $\d R_2/\d t \sim 1/t^2$. 
\textit{Short-range disorder} (the most widespread disorder class, characterized by a finite correlation length, e.g.\ as in the Poissonian objects' placement) 
is represented here by the random packing of non-overlapping spheres. 
Its power spectrum is characterized by a finite plateau at small $k$, ${\overline\Gamma_2(k)|_{k\to 0} = \mathrm{const}}$, such that $p=0$ and $\nu = d/2$. Finally, \textit{strong disorder} is characterized by the diverging structural fluctuations resulting in the exponent $p<0$. An example are randomly placed ``rods", such as vessels or capillaries in the brain (here represented by highly prolate ellipsoids whose long axis exceeds the range of diffusion lengths), yielding $\overline\Gamma_2|_{k\to0} \sim k^{-1}$, $p=-1$, such that $\nu = 1$ and the relaxation rate $R_2(t)\sim \ln t$ diverges.

\begin{figure}[b!]
\centering
\includegraphics[width=\columnwidth]{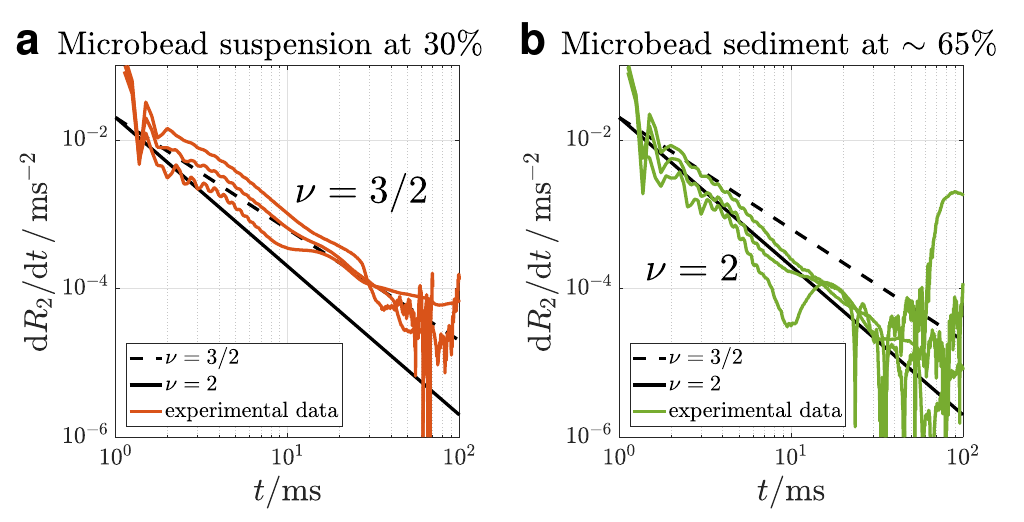}
\vspace{-7mm}
\caption{{\bf Experimental observation of transition into the MRJ state.}
The dynamical exponent (\ref{nu}) changes from $\nu=3/2$ to $\nu=2$ in aqueous suspension of polystyrene microbeads, according to the change of the disorder universality class between the dilute suspension with short-range disorder, $p=0$ (panel {\bf a}), and an MRJ sediment with the predicted exponent\cite{donev2005_PRL} $p=1$ (panel {\bf b}). 
 Shown are representative data sets for water doped with paramagnetic agent ($\mathrm{HoCl_3 \cdot 6H_2O}$) with concentrations
 1.5, 2.0 and $2.5 \units{mmol/L}$ to create a magnetic susceptibility contrast with the microbeads (see Methods for experimental details).
 }
\label{expfig}
\end{figure}

In what follows, we will provide a qualitative coarse-graining argument for the universal relation (\ref{nu}), \fig{fieldfilter}, 
followed by the self-consistent approximation (\ref{dR2pert}) for the signal, \fig{fig:selfconsist}, 
and then demonstrate experimentally how the change of the disorder universality class due to the jamming transition, $p=0 \to p=1$, can be detected via the dynamics (\ref{dR2}) of the measured bulk transverse relaxation, \fig{expfig}. 

An intuition behind the relation of $\d R_2/\d t$ to the spatial fluctuations stems from realizing that the time $t$ defines the diffusion length scale $L(t) = \sqrt{2dDt}$, which acts as a coarse-graining window for the Larmor frequency, 
$\Omega(\r) \to \Omega_t(\r)$, effectively ``seen" by the spins, \fig{fieldfilter}. 
If we were to begin the evolution of magnetization at such $t$ foregoing shorter times, then the coarse-grained $\Omega_t(\r)$ would have the effective correlation length $L(t) > l_c$, and the variance $\la \Omega_t^2(\r)\ra \sim \la \Omega^2(\r)\ra [l_c/L(t)]^d$ decreasing as $ t^{-d/2}$ for the short-range disorder due to Poissonian statistics. 
We now apply the conventional diffusion-narrowing argument $\d R_2 / \d t_c \sim \la \Omega^2\ra$ to our effective $\Omega_t(\r)$, by identifying $t_c \to t$ and $\la \Omega^2\ra \to \la \Omega_t^2\ra$: 
\be \label{RG}
{\d R_{2}(t) \over \d t} \approx \la \Omega^2_t(\r)\ra \sim t^{-\nu}\,,
\ee
with $\nu=d/2$ for the Poissonian case of $p=0$. 

One can view \eq{RG} as a real-space renormalization group equation on the effective macroscopic parameter $R_2$ over the increasing diffusion length scale $L(t)$. We immediately see that the rate $R_2(t)$ {\it always increases} with $t$, as each length scale contributes a strictly positive frequency variance to the relaxation; however, its growth slows down due to the self-averaging, as the instantaneous distribution $\Omega_t(\r)$ becomes narrower with $t$, \fig{fieldfilter}. Moreover, the self-averaging will be faster when the fluctuations $\la \Omega_t^2(\r)\ra$ decrease faster than the inverse ``diffusion volume" $L^{-d}(t)$ (which happens for hyperuniform media,\cite{torquato2003} $p>0$), and slower for strong disorder with diverging fluctuations, $p<0$, in agreement with \fig{fcorrdR2}. 
 
This ``RG flow" must eventually stop for $t > t_\ast$ such that $t_\ast R_2(t_\ast) \sim 1$, when $R_2(t)$ becomes so large that the mesoscopic signal is suppressed exponentially before spins can sample fluctuations of $\Omega(\r)$ at the scales exceeding $L(t_\ast)$.
After $t>t_\ast$, the power law (\ref{dR2}) gets cut-off, \fig{fig:selfconsist}, and by then the mesoscopic signal $s(t)\ll 1$. Hence, the scaling (\ref{dR2}) is detectable for $t_c \ll t \lesssim t_\ast$, provided that the relaxation is sufficiently weak, i.e.\ $R_2(t_c) t_c \ll 1$, which is equivalent to a small ``single-step" phase variance $\la \varphi_1^2\ra \sim \la \Omega^2(\r)\ra t_c^2 \ll 1$. 

The above intuition is supported by finding the disorder-averaged Green's function 
$G_{t,\r-\r_0} = \la \G_{t;\r,\r_0}\ra$ of 
the mesoscopic Bloch-Torrey equation 
for the transverse nuclear magnetization\cite{torrey56,novikov2008,kiselev2018_review} 
\be \label{BT} 
\lb \partial_t - \partial_\r D(\r)\partial_\r + i \Omega(\r) \rb \G_{t;\r,\r_0} = \delta(t) \delta(\r-\r_0) \, ,
\ee 
where $D(\r)$ is the local diffusion coefficient. The mesoscopic contribution $s(t)$ to the NMR signal is helpful to think of in terms of ``spin packets'', the groups of spins emanating from the same point $\r_0$. The magnetization of a spin packet is $\int\!\d\r\, \G_{t;\r,\r_0}$; the $\r_0$- and $t$-dependence of this quantity embodies the coarse-graining discussed above.
Acquisition from a macroscopic sample entails ensemble-averaging of the spin packet magnetization, $s(t) = \frac1{V} \int\! \d\r\d\r_0\, \G_{t;\r,\r_0} \equiv \int\! \d\r\, G_{t,\r}$.
%where $\d\r_0/V$ performs the above averaging of the exact propagator. 
In other words, the signal $s(t)\equiv G_{t,\q}|_{q=0}$ is the Fourier transform of $G_{t,\q}$ for the wavenumber $q=0$. 
Here and in what follows, we factor out the molecular relaxation; the experimentally observable signal $S(t) = e^{-\Rm t} s(t)$. 

We represent the disorder-averaged propagator of \eq{BT} 
\be \label{def-Sigma}
G_{\w,\q} = \frac{1}{-i\w + \Dinf q^2 - \Sigma_{\w,\q}} 
\ee
in terms of the self-energy part\cite{novikov2008,novikov2010} $\Sigma_{\w,\q}$ that collects all one-particle irreducible Feynman diagrams accounting for $\Omega(\r)$ and the deviation $D(\r)- \Dinf$ from the macroscopic diffusion constant $\Dinf$. The expansion of $\Sigma_{\w,\q}$ in the powers of $q$ reflects the measurable mesoscopic effects in the bulk relaxation\cite{novikov2008} (due to even-order correlators $\la \Omega(\r_1) \Omega(\r_2)\dots\ra$), frequency shift (similar correlators of odd orders), diffusion\cite{novikov2010} (due to $\la D(\r_1) D(\r_2)\dots\ra$), and apparent diffusion\cite{novikov2018} (due to the cross-terms $\la \Omega(\r_1) D(\r_2)\dots\ra$). In particular, the mesoscopic spectral lineshape
$s(\w) = \lb -i\w -\Sigma(\w)\rb^{-1}$, where $\Sigma(\w)\equiv \Sigma_{\w,\q}|_{q=0}$. 

We consider the self-energy part in the self-consistent Born approximation,\cite{novikov2008}
equivalent to summing up the ``rainbow" diagrams for $\Sigma_{\w,\q}$:
\be \label{Sigma-sc} 
\Sigma_{\w,\q} \approx - \int\! {\d^d\k \over (2\pi)^d}\, {\Gamma_2(\k)\over -i\w + \Dinf (\q+\k)^2 - \Sigma_{\w,\q+\k} } \,,
\ee 
where the frequency power spectrum, \eq{Gamma_k}, is taken before the angular averaging. The lowest-order approximation to \eq{Sigma-sc} corresponds to neglecting $\Sigma_{\w,\q+\k}$ on the right-hand side, thereby giving the conventional second-order perturbation theory in $\Omega(\r)$\cite{kiselev98,jensen2000_dnr,kiselev2002,sukstanskii2003,sukstanskii2004,novikov2008}, which is asymptotically exact in the diffusion narrowing limit. In this limit, the time domain quantity $\Sigma(t) =  -{\rm d}R_2(t)/{\rm d}t$. 

The next iteration, in the $\w\to0$ limit, is to set $\Sigma_{\w,\q+\k}\to \Sigma(0) \approx -\Rinf$, the terminal relaxation rate, such that
\begin{equation}
\Sigma(\omega) \approx - \int\! \frac{{\rm d}^d\k}{(2 \pi)^d} \, \frac{\overline \Gamma_2(k)}{- i \omega + \Dinf k^2 + \Rinf} \, ,
\label{sigCorrection} 
\end{equation}
equivalent to the temporal scaling
\be\label{dR2pert}
\frac{\rm d}{{\rm d}t} R_2(t) \approx -\Sigma(t) \approx \int\! \frac{\d^dk}{(2 \pi)^d} \, \overline\Gamma_2(k) \ e^{-\Dinf k^2t - \Rinf t} \,,
\ee
where the coarse-grained variance $\la \Omega^2_t(\r)\ra$ is just the Larmor frequency correlation function ``filtered" by the diffusion propagator $e^{-\Dinf k^2t}$, cf.\ \eq{RG}. 
Hence, \fig{fieldfilter} has been obtained using the identification 
$\Omega_t(\k) = \Omega(\k) \, e^{-\Dinf k^2t / 2}$, which represents Gaussian smoothing over the diffusion length $L(t)$. 
Using the low-$k$ scaling (\ref{Gamma_k}), we obtain \eqsand{dR2}{nu} for $t_c\ll t \lesssim t_\ast$. 
The result of leading-order numerical integration (according to \eq{dR2pert} with $\Rinf$ neglected) agrees well with MC simulations in \fig{fcorrdR2} and with asymptotic scaling (\ref{dR2}) for all considered disorder classes. 
The exponent $\nu=1$ for long ellipsoids is equivalent to $\ln s(t) \sim - t \ln t$ for blood vessels.\cite{kiselev98}
For $p=0$, the scaling $\d R_2 /\d t \sim t^{-3/2}$ agrees with the asymptotic behavior for the model medium of Jensen and Chandra\cite{jensen2000_dnr} and for diluted impermeable spheres analyzed by Sukstanskii and Yablonskiy.\cite{sukstanskii2004}

Equivalently, for the leading-order self-energy part, cf.\ \eq{sigCorrection}, we obtain for positive non-integer $\nu$
\begin{equation}
\Sigma(\omega) - \Sigma(0) \sim -\frac{(-i\omega)^{\nu-1}}{\sin\pi\nu}\,,
\label{sigpower} 
\end{equation}
where $-i = e^{-i\pi/2}$. For positiv integer $\nu$, $\Sigma(\omega) - \Sigma(0) \sim (i\omega)^{\nu-1} \ln(-i\omega) $ with account for the regularizing contribution of the region $k\gtrsim k_c$. This gives the specific pattern for $\omega\to 0$ shown in the third row in \fig{fcorrdR2}. These singularities compete with the regular contribution of the domain with finite $k$ in the integral of \eq{sigCorrection} that gives $\Re \Sigma(\w) \sim \omega^2$ and $\Im \Sigma(\w) \sim \omega$. These terms define the form of $\Sigma(\w)$ near $\w=0$ for the regular lattice and dominate $\Im \Sigma(\w)$ for the shuffled lattice ($\nu=5/2$), \fig{fcorrdR2}.

%%%%%%%%%%%%%%%%%%%%%%%
%%%%%%%%%%%%%%%%%%%%%%%

\Fig{fcorrdR2} demonstrates a good agreement between the numerically obtained leading-order singularities and MC simulations. A close look at very small frequencies, $\w t_\ast \lesssim 1$ (\fig{fig:selfconsist}), reveals a lack of accuracy, which is completely cured by the self-consistent Born approximation, \eq{sigCorrection}. According to this equation, the regularization can be viewed as sampling of the leading-oder $\Sigma(\w)$ along the line $\Im \w = \Rinf$, missing the singularity at $\w=0$ in the complex plane of $\w$ by the small value $\Rinf$ (\fig{fig:selfconsist}d). If $\Sigma(\w)$ is measured without correcting for the molecular relaxation, this shift increases by the substitution $\Rinf \to R_2^{\rm mol}+\Rinf$. 

The present discussion shows that the universality, our main result expressed by \eqsand{dR2}{nu}, is mapped onto the behavior of the structural power spectrum for small $k$, which is independent of individual properties of the magnetized objects. This is illustrated in supplementary \fig{imperfigs} that shows results of MC simulations for the same media as in \fig{fcorrdR2}, but with the magnetized objects made {\it impermeable} for diffusing spins. 

As an application of the developed formalism, in \fig{expfig} we experimentally demonstrate the change of the disorder universality class $p=0\to 1$ after reaching the maximally random jammed state for mono-dispersed spheres, where the nontrivial exponent $p=1$ was predicted numerically.\cite{donev2005_PRL}
NMR relaxation in two microbead packings (suspension and densely-packed sediment) exhibit distinct exponents for the time derivative of the relaxation rate, which makes it possible to distinguish the two packings using a {\it macroscopic} NMR measurement in contrast to the microsopic character of up-to-date observations.\cite{Xie2013,Jiao2014,Dreyfus2015} 
This remarkable sensitivity of the macroscopic measurement to the nontrivial mesoscopic structure is enabled by the time-dependent coarse-graining window, Fig.~\ref{fieldfilter}, that effectively samples the mesoscopic medium's power spectrum (\ref{Gamma_chi}).

To conclude, we have shown analytically and numerically, that the mesoscopic component of the transverse relaxation rate displays a universal scaling behavior that is sensitive to the statistics of large-scale organization of tissue magnetic susceptibility. This allowed us to provide the first macroscopic experimental observation of the MRJ transition in spherical microbead packings using NMR relaxation. 
Our results provide a framework for noninvasive investigation of the structure of complex materials and in biomedical magnetic resonance imaging, where both native and added susceptibility contrast is ubiquitous.

\bibliography{library}

\begin{thebibliography}{38}%
\makeatletter
\providecommand \@ifxundefined [1]{%
 \@ifx{#1\undefined}
}%
\providecommand \@ifnum [1]{%
 \ifnum #1\expandafter \@firstoftwo
 \else \expandafter \@secondoftwo
 \fi
}%
\providecommand \@ifx [1]{%
 \ifx #1\expandafter \@firstoftwo
 \else \expandafter \@secondoftwo
 \fi
}%
\providecommand \natexlab [1]{#1}%
\providecommand \enquote  [1]{``#1''}%
\providecommand \bibnamefont  [1]{#1}%
\providecommand \bibfnamefont [1]{#1}%
\providecommand \citenamefont [1]{#1}%
\providecommand \href@noop [0]{\@secondoftwo}%
\providecommand \href [0]{\begingroup \@sanitize@url \@href}%
\providecommand \@href[1]{\@@startlink{#1}\@@href}%
\providecommand \@@href[1]{\endgroup#1\@@endlink}%
\providecommand \@sanitize@url [0]{\catcode `\\12\catcode `\$12\catcode
  `\&12\catcode `\#12\catcode `\^12\catcode `\_12\catcode `\%12\relax}%
\providecommand \@@startlink[1]{}%
\providecommand \@@endlink[0]{}%
\providecommand \url  [0]{\begingroup\@sanitize@url \@url }%
\providecommand \@url [1]{\endgroup\@href {#1}{\urlprefix }}%
\providecommand \urlprefix  [0]{URL }%
\providecommand \Eprint [0]{\href }%
\providecommand \doibase [0]{http://dx.doi.org/}%
\providecommand \selectlanguage [0]{\@gobble}%
\providecommand \bibinfo  [0]{\@secondoftwo}%
\providecommand \bibfield  [0]{\@secondoftwo}%
\providecommand \translation [1]{[#1]}%
\providecommand \BibitemOpen [0]{}%
\providecommand \bibitemStop [0]{}%
\providecommand \bibitemNoStop [0]{.\EOS\space}%
\providecommand \EOS [0]{\spacefactor3000\relax}%
\providecommand \BibitemShut  [1]{\csname bibitem#1\endcsname}%
\let\auto@bib@innerbib\@empty
%</preamble>
\bibitem [{\citenamefont {Bloembergen}\ \emph {et~al.}(1948)\citenamefont
  {Bloembergen}, \citenamefont {Purcell},\ and\ \citenamefont
  {Pound}}]{bloembergen48}%
  \BibitemOpen
  \bibfield  {author} {\bibinfo {author} {\bibfnamefont {N.}~\bibnamefont
  {Bloembergen}}, \bibinfo {author} {\bibfnamefont {E.~M.}\ \bibnamefont
  {Purcell}}, \ and\ \bibinfo {author} {\bibfnamefont {R.~V.}\ \bibnamefont
  {Pound}},\ }\bibfield  {title} {\enquote {\bibinfo {title} {{Relaxation
  Effects in Nuclear Magnetic Resonance Absorption}},}\ }\href {\doibase
  10.1103/PhysRev.73.679} {\bibfield  {journal} {\bibinfo  {journal} {Phys
  Rev}\ }\textbf {\bibinfo {volume} {73}},\ \bibinfo {pages} {679--712}
  (\bibinfo {year} {1948})}\BibitemShut {NoStop}%
\bibitem [{\citenamefont {Anderson}\ and\ \citenamefont
  {Weiss}(1953)}]{anderson53}%
  \BibitemOpen
  \bibfield  {author} {\bibinfo {author} {\bibfnamefont {P.~W.}\ \bibnamefont
  {Anderson}}\ and\ \bibinfo {author} {\bibfnamefont {P.~R.}\ \bibnamefont
  {Weiss}},\ }\bibfield  {title} {\enquote {\bibinfo {title} {{Exchange
  Narrowing in Paramagnetic Resonance}},}\ }\href {\doibase
  10.1103/RevModPhys.25.269} {\bibfield  {journal} {\bibinfo  {journal} {Rev
  Mod Phys}\ }\textbf {\bibinfo {volume} {25}},\ \bibinfo {pages} {269--276}
  (\bibinfo {year} {1953})}\BibitemShut {NoStop}%
\bibitem [{\citenamefont {Khaetskii}\ \emph {et~al.}(2002)\citenamefont
  {Khaetskii}, \citenamefont {Loss},\ and\ \citenamefont
  {Glazman}}]{Khaetskii2002}%
  \BibitemOpen
  \bibfield  {author} {\bibinfo {author} {\bibfnamefont {A.~V.}\ \bibnamefont
  {Khaetskii}}, \bibinfo {author} {\bibfnamefont {D.}~\bibnamefont {Loss}}, \
  and\ \bibinfo {author} {\bibfnamefont {L.}~\bibnamefont {Glazman}},\
  }\bibfield  {title} {\enquote {\bibinfo {title} {{Electron Spin Decoherence
  in Quantum Dots due to Interaction with Nuclei}},}\ }\href {\doibase
  10.1103/PhysRevLett.88.186802} {\bibfield  {journal} {\bibinfo  {journal}
  {Phys Rev Lett}\ }\textbf {\bibinfo {volume} {88}},\ \bibinfo {pages}
  {186802} (\bibinfo {year} {2002})}\BibitemShut {NoStop}%
\bibitem [{\citenamefont {Uhrig}(2007)}]{Uhrig2007}%
  \BibitemOpen
  \bibfield  {author} {\bibinfo {author} {\bibfnamefont {G.~S.}\ \bibnamefont
  {Uhrig}},\ }\bibfield  {title} {\enquote {\bibinfo {title} {{Keeping a
  Quantum Bit Alive by Optimized $\pi$-Pulse Sequences}},}\ }\href {\doibase
  10.1103/PhysRevLett.98.100504} {\bibfield  {journal} {\bibinfo  {journal}
  {Phys Rev Lett}\ }\textbf {\bibinfo {volume} {98}},\ \bibinfo {pages}
  {100504} (\bibinfo {year} {2007})}\BibitemShut {NoStop}%
\bibitem [{\citenamefont {D'yakonov}\ and\ \citenamefont
  {Perel'}(1971)}]{Dyakonov71}%
  \BibitemOpen
  \bibfield  {author} {\bibinfo {author} {\bibfnamefont {M.~I.}\ \bibnamefont
  {D'yakonov}}\ and\ \bibinfo {author} {\bibfnamefont {V.~I.}\ \bibnamefont
  {Perel'}},\ }\bibfield  {title} {\enquote {\bibinfo {title} {{Spin
  orientation of electrons associated with the interband absorption of light in
  semiconductors}},}\ }\href@noop {} {\bibfield  {journal} {\bibinfo  {journal}
  {Sov Phys JETP}\ }\textbf {\bibinfo {volume} {33}},\ \bibinfo {pages}
  {1053--1059} (\bibinfo {year} {1971})}\BibitemShut {NoStop}%
\bibitem [{\citenamefont {Salis}\ \emph {et~al.}(2001)\citenamefont {Salis},
  \citenamefont {Kato}, \citenamefont {Ensslin}, \citenamefont {Driscoll},
  \citenamefont {Gossard},\ and\ \citenamefont {Awschalom}}]{Salis2001}%
  \BibitemOpen
  \bibfield  {author} {\bibinfo {author} {\bibfnamefont {G.}~\bibnamefont
  {Salis}}, \bibinfo {author} {\bibfnamefont {Y.}~\bibnamefont {Kato}},
  \bibinfo {author} {\bibfnamefont {K.}~\bibnamefont {Ensslin}}, \bibinfo
  {author} {\bibfnamefont {D.~C.}\ \bibnamefont {Driscoll}}, \bibinfo {author}
  {\bibfnamefont {A.~C.}\ \bibnamefont {Gossard}}, \ and\ \bibinfo {author}
  {\bibfnamefont {D.~D.}\ \bibnamefont {Awschalom}},\ }\bibfield  {title}
  {\enquote {\bibinfo {title} {{Electrical control of spin coherence in
  semiconductor nanostructures}},}\ }\href {\doibase 10.1038/414619a}
  {\bibfield  {journal} {\bibinfo  {journal} {Nature}\ }\textbf {\bibinfo
  {volume} {414}},\ \bibinfo {pages} {619--622} (\bibinfo {year}
  {2001})}\BibitemShut {NoStop}%
\bibitem [{\citenamefont {H{\"{u}}rlimann}(1998)}]{hurlimann98}%
  \BibitemOpen
  \bibfield  {author} {\bibinfo {author} {\bibfnamefont {M.~D.}\ \bibnamefont
  {H{\"{u}}rlimann}},\ }\bibfield  {title} {\enquote {\bibinfo {title}
  {{Effective Gradients in Porous Media Due to Susceptibility Differences}},}\
  }\href {\doibase 10.1006/jmre.1998.1364} {\bibfield  {journal} {\bibinfo
  {journal} {J Magn Reson}\ }\textbf {\bibinfo {volume} {131}},\ \bibinfo
  {pages} {232--240} (\bibinfo {year} {1998})}\BibitemShut {NoStop}%
\bibitem [{\citenamefont {Song}\ \emph {et~al.}(2000)\citenamefont {Song},
  \citenamefont {Ryu},\ and\ \citenamefont {Sen}}]{song2000}%
  \BibitemOpen
  \bibfield  {author} {\bibinfo {author} {\bibfnamefont {Y.-Q.}\ \bibnamefont
  {Song}}, \bibinfo {author} {\bibfnamefont {S.}~\bibnamefont {Ryu}}, \ and\
  \bibinfo {author} {\bibfnamefont {P.~N.}\ \bibnamefont {Sen}},\ }\bibfield
  {title} {\enquote {\bibinfo {title} {{Determining multiple length scales in
  rocks}},}\ }\href {\doibase 10.1038/35018057} {\bibfield  {journal} {\bibinfo
   {journal} {Nature}\ }\textbf {\bibinfo {volume} {406}},\ \bibinfo {pages}
  {178--181} (\bibinfo {year} {2000})}\BibitemShut {NoStop}%
\bibitem [{\citenamefont {Glasel}\ and\ \citenamefont {Lee}(1974)}]{glasel74}%
  \BibitemOpen
  \bibfield  {author} {\bibinfo {author} {\bibfnamefont {J.~A.}\ \bibnamefont
  {Glasel}}\ and\ \bibinfo {author} {\bibfnamefont {K.~H.}\ \bibnamefont
  {Lee}},\ }\bibfield  {title} {\enquote {\bibinfo {title} {{On the
  Interpretation of Water Nuclear Magnetic Resonance Relaxation Times in
  Heterogeneous Systems}},}\ }\href {\doibase 10.1021/ja00811a003} {\bibfield
  {journal} {\bibinfo  {journal} {J Am Chem Soc}\ }\textbf {\bibinfo {volume}
  {96}},\ \bibinfo {pages} {970--978} (\bibinfo {year} {1974})}\BibitemShut
  {NoStop}%
\bibitem [{\citenamefont {Thulborn}\ \emph {et~al.}(1982)\citenamefont
  {Thulborn}, \citenamefont {Waterton}, \citenamefont {Matthews},\ and\
  \citenamefont {Radda}}]{thulborn82}%
  \BibitemOpen
  \bibfield  {author} {\bibinfo {author} {\bibfnamefont {K.~R.}\ \bibnamefont
  {Thulborn}}, \bibinfo {author} {\bibfnamefont {J.~C.}\ \bibnamefont
  {Waterton}}, \bibinfo {author} {\bibfnamefont {P.~M.}\ \bibnamefont
  {Matthews}}, \ and\ \bibinfo {author} {\bibfnamefont {G.~K.}\ \bibnamefont
  {Radda}},\ }\bibfield  {title} {\enquote {\bibinfo {title} {{Oxygenation
  dependence of the transverse relaxation time of water protons in whole blood
  at high field}},}\ }\href {\doibase 10.1016/0304-4165(82)90333-6} {\bibfield
  {journal} {\bibinfo  {journal} {Biochim Biophys Acta}\ }\textbf {\bibinfo
  {volume} {714}},\ \bibinfo {pages} {265--270} (\bibinfo {year}
  {1982})}\BibitemShut {NoStop}%
\bibitem [{\citenamefont {Gillis}\ and\ \citenamefont
  {Koenig}(1987)}]{gillis87}%
  \BibitemOpen
  \bibfield  {author} {\bibinfo {author} {\bibfnamefont {P.}~\bibnamefont
  {Gillis}}\ and\ \bibinfo {author} {\bibfnamefont {S.~H.}\ \bibnamefont
  {Koenig}},\ }\bibfield  {title} {\enquote {\bibinfo {title} {{Transverse
  Relaxation of Solvent Protons Induced by Magnetized Spheres: Application to
  Ferritin, Erythrocytes, and Magnetite}},}\ }\href {\doibase
  10.1002/mrm.1910050404} {\bibfield  {journal} {\bibinfo  {journal} {Magn
  Reson Med}\ }\textbf {\bibinfo {volume} {5}},\ \bibinfo {pages} {323--345}
  (\bibinfo {year} {1987})}\BibitemShut {NoStop}%
\bibitem [{\citenamefont {Weisskoff}\ \emph {et~al.}(1994)\citenamefont
  {Weisskoff}, \citenamefont {Zuo}, \citenamefont {Boxerman},\ and\
  \citenamefont {Rosen}}]{Weisskoff94}%
  \BibitemOpen
  \bibfield  {author} {\bibinfo {author} {\bibfnamefont {R.~M.}\ \bibnamefont
  {Weisskoff}}, \bibinfo {author} {\bibfnamefont {C.~S.}\ \bibnamefont {Zuo}},
  \bibinfo {author} {\bibfnamefont {J.~L.}\ \bibnamefont {Boxerman}}, \ and\
  \bibinfo {author} {\bibfnamefont {B.~R.}\ \bibnamefont {Rosen}},\ }\bibfield
  {title} {\enquote {\bibinfo {title} {{Microscopic Susceptibility Variation
  and Transverse Relaxation: Theory and Experiment}},}\ }\href {\doibase
  10.1002/mrm.1910310605} {\bibfield  {journal} {\bibinfo  {journal} {Magn
  Reson Med}\ }\textbf {\bibinfo {volume} {31}},\ \bibinfo {pages} {601--610}
  (\bibinfo {year} {1994})}\BibitemShut {NoStop}%
\bibitem [{\citenamefont {Davis}\ \emph {et~al.}(2018)\citenamefont {Davis},
  \citenamefont {Ramesh}, \citenamefont {Bhatnagar}, \citenamefont
  {Lee-Gosselin}, \citenamefont {Barry}, \citenamefont {Glenn}, \citenamefont
  {Walsworth},\ and\ \citenamefont {Shapiro}}]{Davis2018}%
  \BibitemOpen
  \bibfield  {author} {\bibinfo {author} {\bibfnamefont {H.~C.}\ \bibnamefont
  {Davis}}, \bibinfo {author} {\bibfnamefont {P.}~\bibnamefont {Ramesh}},
  \bibinfo {author} {\bibfnamefont {A.}~\bibnamefont {Bhatnagar}}, \bibinfo
  {author} {\bibfnamefont {A.}~\bibnamefont {Lee-Gosselin}}, \bibinfo {author}
  {\bibfnamefont {J.~F.}\ \bibnamefont {Barry}}, \bibinfo {author}
  {\bibfnamefont {D.~R.}\ \bibnamefont {Glenn}}, \bibinfo {author}
  {\bibfnamefont {R.~L.}\ \bibnamefont {Walsworth}}, \ and\ \bibinfo {author}
  {\bibfnamefont {M.~G.}\ \bibnamefont {Shapiro}},\ }\bibfield  {title}
  {\enquote {\bibinfo {title} {{Mapping the microscale origins of magnetic
  resonance image contrast with subcellular diamond magnetometry}},}\ }\href
  {\doibase 10.1038/s41467-017-02471-7} {\bibfield  {journal} {\bibinfo
  {journal} {Nat Commun}\ }\textbf {\bibinfo {volume} {9}},\ \bibinfo {pages}
  {131} (\bibinfo {year} {2018})}\BibitemShut {NoStop}%
\bibitem [{\citenamefont {Kiselev}\ and\ \citenamefont
  {Novikov}(2018)}]{kiselev2018_review}%
  \BibitemOpen
  \bibfield  {author} {\bibinfo {author} {\bibfnamefont {V.~G.}\ \bibnamefont
  {Kiselev}}\ and\ \bibinfo {author} {\bibfnamefont {D.~S.}\ \bibnamefont
  {Novikov}},\ }\bibfield  {title} {\enquote {\bibinfo {title} {{Transverse NMR
  relaxation in biological tissues}},}\ }\href {\doibase
  10.1016/j.neuroimage.2018.06.002} {\bibfield  {journal} {\bibinfo  {journal}
  {NeuroImage}\ } (\bibinfo {year} {2018}),\
  10.1016/j.neuroimage.2018.06.002}\BibitemShut {NoStop}%
\bibitem [{\citenamefont {Yablonskiy}\ and\ \citenamefont
  {Haacke}(1994)}]{yablonskiy94}%
  \BibitemOpen
  \bibfield  {author} {\bibinfo {author} {\bibfnamefont {D.~A.}\ \bibnamefont
  {Yablonskiy}}\ and\ \bibinfo {author} {\bibfnamefont {E.~M.}\ \bibnamefont
  {Haacke}},\ }\bibfield  {title} {\enquote {\bibinfo {title} {{Theory of {NMR}
  Signal Behavior in Magnetically Inhomogeneous Tissues: {T}he Static Dephasing
  Regime}},}\ }\href {\doibase 10.1002/mrm.1910320610} {\bibfield  {journal}
  {\bibinfo  {journal} {Magn Reson Med}\ }\textbf {\bibinfo {volume} {32}},\
  \bibinfo {pages} {749--763} (\bibinfo {year} {1994})}\BibitemShut {NoStop}%
\bibitem [{\citenamefont {Kiselev}\ and\ \citenamefont
  {Posse}(1998)}]{kiselev98}%
  \BibitemOpen
  \bibfield  {author} {\bibinfo {author} {\bibfnamefont {V.~G.}\ \bibnamefont
  {Kiselev}}\ and\ \bibinfo {author} {\bibfnamefont {S.}~\bibnamefont
  {Posse}},\ }\bibfield  {title} {\enquote {\bibinfo {title} {{Analytical
  Theory of Susceptibility Induced NMR Signal Dephasing in a Cerebrovascular
  Network}},}\ }\href {\doibase 10.1103/PhysRevLett.81.5696} {\bibfield
  {journal} {\bibinfo  {journal} {Phys Rev Lett}\ }\textbf {\bibinfo {volume}
  {81}},\ \bibinfo {pages} {5696--5699} (\bibinfo {year} {1998})}\BibitemShut
  {NoStop}%
\bibitem [{\citenamefont {Jensen}\ and\ \citenamefont
  {Chandra}(2000)}]{jensen2000_dnr}%
  \BibitemOpen
  \bibfield  {author} {\bibinfo {author} {\bibfnamefont {J.~H.}\ \bibnamefont
  {Jensen}}\ and\ \bibinfo {author} {\bibfnamefont {R.}~\bibnamefont
  {Chandra}},\ }\bibfield  {title} {\enquote {\bibinfo {title} {{NMR Relaxation
  in Tissues With Weak Magnetic Inhomogeneities}},}\ }\href {\doibase
  10.1002/1522-2594(200007)44:1<144::AID-MRM21>3.0.CO;2-O} {\bibfield
  {journal} {\bibinfo  {journal} {Magn Reson Med}\ }\textbf {\bibinfo {volume}
  {44}},\ \bibinfo {pages} {144--156} (\bibinfo {year} {2000})}\BibitemShut
  {NoStop}%
\bibitem [{\citenamefont {Kiselev}\ and\ \citenamefont
  {Novikov}(2002)}]{kiselev2002}%
  \BibitemOpen
  \bibfield  {author} {\bibinfo {author} {\bibfnamefont {V.~G.}\ \bibnamefont
  {Kiselev}}\ and\ \bibinfo {author} {\bibfnamefont {D.~S.}\ \bibnamefont
  {Novikov}},\ }\bibfield  {title} {\enquote {\bibinfo {title} {{Transverse
  {NMR} Relaxation as a Probe of Mesoscopic Structure}},}\ }\href {\doibase
  10.1103/PhysRevLett.89.278101} {\bibfield  {journal} {\bibinfo  {journal}
  {Phys Rev Lett}\ }\textbf {\bibinfo {volume} {89}},\ \bibinfo {pages}
  {278101} (\bibinfo {year} {2002})}\BibitemShut {NoStop}%
\bibitem [{\citenamefont {Sukstanskii}\ and\ \citenamefont
  {Yablonskiy}(2003)}]{sukstanskii2003}%
  \BibitemOpen
  \bibfield  {author} {\bibinfo {author} {\bibfnamefont {A.~L.}\ \bibnamefont
  {Sukstanskii}}\ and\ \bibinfo {author} {\bibfnamefont {D.~A.}\ \bibnamefont
  {Yablonskiy}},\ }\bibfield  {title} {\enquote {\bibinfo {title} {{Gaussian
  approximation in the theory of MR signal formation in the presence of
  structure-specific magnetic field inhomogeneities}},}\ }\href {\doibase
  10.1016/S1090-7807(03)00131-9} {\bibfield  {journal} {\bibinfo  {journal} {J
  Magn Reson}\ }\textbf {\bibinfo {volume} {163}},\ \bibinfo {pages} {236--247}
  (\bibinfo {year} {2003})}\BibitemShut {NoStop}%
\bibitem [{\citenamefont {Sukstanskii}\ and\ \citenamefont
  {Yablonskiy}(2004)}]{sukstanskii2004}%
  \BibitemOpen
  \bibfield  {author} {\bibinfo {author} {\bibfnamefont {A.~L.}\ \bibnamefont
  {Sukstanskii}}\ and\ \bibinfo {author} {\bibfnamefont {D.~A.}\ \bibnamefont
  {Yablonskiy}},\ }\bibfield  {title} {\enquote {\bibinfo {title} {{Gaussian
  approximation in the theory of MR signal formation in the presence of
  structure-specific magnetic field inhomogeneities. Effects of impermeable
  susceptibility inclusions}},}\ }\href {\doibase 10.1016/j.jmr.2003.11.006}
  {\bibfield  {journal} {\bibinfo  {journal} {J Magn Reson}\ }\textbf {\bibinfo
  {volume} {167}},\ \bibinfo {pages} {56--67} (\bibinfo {year}
  {2004})}\BibitemShut {NoStop}%
\bibitem [{\citenamefont {Novikov}\ and\ \citenamefont
  {Kiselev}(2008)}]{novikov2008}%
  \BibitemOpen
  \bibfield  {author} {\bibinfo {author} {\bibfnamefont {D.~S.}\ \bibnamefont
  {Novikov}}\ and\ \bibinfo {author} {\bibfnamefont {V.~G.}\ \bibnamefont
  {Kiselev}},\ }\bibfield  {title} {\enquote {\bibinfo {title} {{Transverse
  {NMR} relaxation in magnetically heterogeneous media}},}\ }\href {\doibase
  10.1016/j.jmr.2008.08.005} {\bibfield  {journal} {\bibinfo  {journal} {J Magn
  Reson}\ }\textbf {\bibinfo {volume} {195}},\ \bibinfo {pages} {33--39}
  (\bibinfo {year} {2008})}\BibitemShut {NoStop}%
\bibitem [{\citenamefont {Novikov}\ \emph {et~al.}(2014)\citenamefont
  {Novikov}, \citenamefont {Jensen}, \citenamefont {Helpern},\ and\
  \citenamefont {Fieremans}}]{novikov2014}%
  \BibitemOpen
  \bibfield  {author} {\bibinfo {author} {\bibfnamefont {D.~S.}\ \bibnamefont
  {Novikov}}, \bibinfo {author} {\bibfnamefont {J.~H.}\ \bibnamefont {Jensen}},
  \bibinfo {author} {\bibfnamefont {J.~A.}\ \bibnamefont {Helpern}}, \ and\
  \bibinfo {author} {\bibfnamefont {E.}~\bibnamefont {Fieremans}},\ }\bibfield
  {title} {\enquote {\bibinfo {title} {{Revealing mesoscopic structural
  universality with diffusion}},}\ }\href {\doibase 10.1073/pnas.1316944111}
  {\bibfield  {journal} {\bibinfo  {journal} {Proc Natl Acad Sci USA}\ }\textbf
  {\bibinfo {volume} {111}},\ \bibinfo {pages} {5088--5093} (\bibinfo {year}
  {2014})}\BibitemShut {NoStop}%
\bibitem [{\citenamefont {Torquato}\ \emph {et~al.}(2000)\citenamefont
  {Torquato}, \citenamefont {Truskett},\ and\ \citenamefont
  {Debenedetti}}]{torquato2000}%
  \BibitemOpen
  \bibfield  {author} {\bibinfo {author} {\bibfnamefont {S.}~\bibnamefont
  {Torquato}}, \bibinfo {author} {\bibfnamefont {T.~M.}\ \bibnamefont
  {Truskett}}, \ and\ \bibinfo {author} {\bibfnamefont {P.~G.}\ \bibnamefont
  {Debenedetti}},\ }\bibfield  {title} {\enquote {\bibinfo {title} {{Is Random
  Close Packing of Spheres Well Defined?}}}\ }\href {\doibase
  10.1103/PhysRevLett.84.2064} {\bibfield  {journal} {\bibinfo  {journal} {Phys
  Rev Lett}\ }\textbf {\bibinfo {volume} {84}},\ \bibinfo {pages} {2064--2067}
  (\bibinfo {year} {2000})}\BibitemShut {NoStop}%
\bibitem [{\citenamefont {Torquato}\ and\ \citenamefont
  {Stillinger}(2003)}]{torquato2003}%
  \BibitemOpen
  \bibfield  {author} {\bibinfo {author} {\bibfnamefont {S.}~\bibnamefont
  {Torquato}}\ and\ \bibinfo {author} {\bibfnamefont {F.~H.}\ \bibnamefont
  {Stillinger}},\ }\bibfield  {title} {\enquote {\bibinfo {title} {{Local
  density fluctuations, hyperuniformity, and order metrics}},}\ }\href
  {\doibase 10.1103/PhysRevE.68.041113} {\bibfield  {journal} {\bibinfo
  {journal} {Phys Rev E}\ }\textbf {\bibinfo {volume} {68}},\ \bibinfo {pages}
  {41113} (\bibinfo {year} {2003})}\BibitemShut {NoStop}%
\bibitem [{\citenamefont {Donev}\ \emph {et~al.}(2005)\citenamefont {Donev},
  \citenamefont {Stillinger},\ and\ \citenamefont {Torquato}}]{donev2005_PRL}%
  \BibitemOpen
  \bibfield  {author} {\bibinfo {author} {\bibfnamefont {A.}~\bibnamefont
  {Donev}}, \bibinfo {author} {\bibfnamefont {F.~H.}\ \bibnamefont
  {Stillinger}}, \ and\ \bibinfo {author} {\bibfnamefont {S.}~\bibnamefont
  {Torquato}},\ }\bibfield  {title} {\enquote {\bibinfo {title} {{Unexpected
  Density Fluctuations in Jammed Disordered Sphere Packings}},}\ }\href
  {\doibase 10.1103/PhysRevLett.95.090604} {\bibfield  {journal} {\bibinfo
  {journal} {Phys Rev Lett}\ }\textbf {\bibinfo {volume} {95}},\ \bibinfo
  {pages} {90604} (\bibinfo {year} {2005})}\BibitemShut {NoStop}%
\bibitem [{\citenamefont {Kennan}\ \emph {et~al.}(1994)\citenamefont {Kennan},
  \citenamefont {Zhong},\ and\ \citenamefont {Gore}}]{kennan94}%
  \BibitemOpen
  \bibfield  {author} {\bibinfo {author} {\bibfnamefont {R.~P.}\ \bibnamefont
  {Kennan}}, \bibinfo {author} {\bibfnamefont {J.}~\bibnamefont {Zhong}}, \
  and\ \bibinfo {author} {\bibfnamefont {J.~C.}\ \bibnamefont {Gore}},\
  }\bibfield  {title} {\enquote {\bibinfo {title} {{Intravascular
  Susceptibility Contrast Mechanisms in Tissues}},}\ }\href {\doibase
  10.1002/mrm.1910310103} {\bibfield  {journal} {\bibinfo  {journal} {Magn
  Reson Med}\ }\textbf {\bibinfo {volume} {31}},\ \bibinfo {pages} {9--21}
  (\bibinfo {year} {1994})}\BibitemShut {NoStop}%
\bibitem [{\citenamefont {Gillis}\ \emph {et~al.}(1995)\citenamefont {Gillis},
  \citenamefont {Pet{\"{o}}}, \citenamefont {Moiny}, \citenamefont
  {Mispelter},\ and\ \citenamefont {Cuenod}}]{gillis95}%
  \BibitemOpen
  \bibfield  {author} {\bibinfo {author} {\bibfnamefont {P.}~\bibnamefont
  {Gillis}}, \bibinfo {author} {\bibfnamefont {S.}~\bibnamefont {Pet{\"{o}}}},
  \bibinfo {author} {\bibfnamefont {F.}~\bibnamefont {Moiny}}, \bibinfo
  {author} {\bibfnamefont {J.}~\bibnamefont {Mispelter}}, \ and\ \bibinfo
  {author} {\bibfnamefont {C.~A.}\ \bibnamefont {Cuenod}},\ }\bibfield  {title}
  {\enquote {\bibinfo {title} {{Proton Transverse Nuclear Magnetic Relaxation
  in Oxidized Blood: a Numerical Approach}},}\ }\href {\doibase
  10.1002/mrm.1910330114} {\bibfield  {journal} {\bibinfo  {journal} {Magn
  Reson Med}\ }\textbf {\bibinfo {volume} {33}},\ \bibinfo {pages} {93--100}
  (\bibinfo {year} {1995})}\BibitemShut {NoStop}%
\bibitem [{\citenamefont {Deville}\ \emph {et~al.}(1979)\citenamefont
  {Deville}, \citenamefont {Bernier},\ and\ \citenamefont
  {Delrieux}}]{deville79}%
  \BibitemOpen
  \bibfield  {author} {\bibinfo {author} {\bibfnamefont {G.}~\bibnamefont
  {Deville}}, \bibinfo {author} {\bibfnamefont {M.}~\bibnamefont {Bernier}}, \
  and\ \bibinfo {author} {\bibfnamefont {J.~M.}\ \bibnamefont {Delrieux}},\
  }\bibfield  {title} {\enquote {\bibinfo {title} {{NMR multiple echoes
  observed in solid $^{3}\mathrm{He}$}},}\ }\href {\doibase
  10.1103/PhysRevB.19.5666} {\bibfield  {journal} {\bibinfo  {journal} {Phys
  Rev B}\ }\textbf {\bibinfo {volume} {19}},\ \bibinfo {pages} {5666--5688}
  (\bibinfo {year} {1979})}\BibitemShut {NoStop}%
\bibitem [{\citenamefont {Gabrielli}\ \emph {et~al.}(2002)\citenamefont
  {Gabrielli}, \citenamefont {Joyce},\ and\ \citenamefont {{Sylos
  Labini}}}]{gabrielli2002}%
  \BibitemOpen
  \bibfield  {author} {\bibinfo {author} {\bibfnamefont {A.}~\bibnamefont
  {Gabrielli}}, \bibinfo {author} {\bibfnamefont {M.}~\bibnamefont {Joyce}}, \
  and\ \bibinfo {author} {\bibfnamefont {F.}~\bibnamefont {{Sylos Labini}}},\
  }\bibfield  {title} {\enquote {\bibinfo {title} {{Glass-like universe:
  Real-space correlation properties of standard cosmological models}},}\ }\href
  {\doibase 10.1103/PhysRevD.65.083523} {\bibfield  {journal} {\bibinfo
  {journal} {Phys Rev D}\ }\textbf {\bibinfo {volume} {65}},\ \bibinfo {pages}
  {83523} (\bibinfo {year} {2002})}\BibitemShut {NoStop}%
\bibitem [{\citenamefont {Torrey}(1956)}]{torrey56}%
  \BibitemOpen
  \bibfield  {author} {\bibinfo {author} {\bibfnamefont {H.~C.}\ \bibnamefont
  {Torrey}},\ }\bibfield  {title} {\enquote {\bibinfo {title} {{Bloch Equations
  with Diffusion Terms}},}\ }\href {\doibase 10.1103/PhysRev.104.563}
  {\bibfield  {journal} {\bibinfo  {journal} {Phys Rev}\ }\textbf {\bibinfo
  {volume} {104}},\ \bibinfo {pages} {563--565} (\bibinfo {year}
  {1956})}\BibitemShut {NoStop}%
\bibitem [{\citenamefont {Novikov}\ and\ \citenamefont
  {Kiselev}(2010)}]{novikov2010}%
  \BibitemOpen
  \bibfield  {author} {\bibinfo {author} {\bibfnamefont {D.~S.}\ \bibnamefont
  {Novikov}}\ and\ \bibinfo {author} {\bibfnamefont {V.~G.}\ \bibnamefont
  {Kiselev}},\ }\bibfield  {title} {\enquote {\bibinfo {title} {{Effective
  medium theory of a diffusion-weighted signal}},}\ }\href {\doibase
  10.1002/nbm.1584} {\bibfield  {journal} {\bibinfo  {journal} {NMR Biomed}\
  }\textbf {\bibinfo {volume} {23}},\ \bibinfo {pages} {682--697} (\bibinfo
  {year} {2010})}\BibitemShut {NoStop}%
\bibitem [{\citenamefont {Novikov}\ \emph {et~al.}(2018)\citenamefont
  {Novikov}, \citenamefont {Reisert},\ and\ \citenamefont
  {Kiselev}}]{novikov2018}%
  \BibitemOpen
  \bibfield  {author} {\bibinfo {author} {\bibfnamefont {D.~S.}\ \bibnamefont
  {Novikov}}, \bibinfo {author} {\bibfnamefont {M.}~\bibnamefont {Reisert}}, \
  and\ \bibinfo {author} {\bibfnamefont {V.~G.}\ \bibnamefont {Kiselev}},\
  }\bibfield  {title} {\enquote {\bibinfo {title} {{Effects of mesoscopic
  susceptibility and transverse relaxation on diffusion NMR}},}\ }\href
  {\doibase 10.1016/j.jmr.2018.06.007} {\bibfield  {journal} {\bibinfo
  {journal} {J Magn Reson}\ }\textbf {\bibinfo {volume} {293}},\ \bibinfo
  {pages} {134--144} (\bibinfo {year} {2018})}\BibitemShut {NoStop}%
\bibitem [{\citenamefont {Xie}\ \emph {et~al.}(2013)\citenamefont {Xie},
  \citenamefont {Long}, \citenamefont {Weigand}, \citenamefont {Moss},
  \citenamefont {Carvalho}, \citenamefont {Roorda}, \citenamefont {Hejna},
  \citenamefont {Torquato},\ and\ \citenamefont {Steinhardt}}]{Xie2013}%
  \BibitemOpen
  \bibfield  {author} {\bibinfo {author} {\bibfnamefont {R.}~\bibnamefont
  {Xie}}, \bibinfo {author} {\bibfnamefont {G.~G.}\ \bibnamefont {Long}},
  \bibinfo {author} {\bibfnamefont {S.~J.}\ \bibnamefont {Weigand}}, \bibinfo
  {author} {\bibfnamefont {S.~C.}\ \bibnamefont {Moss}}, \bibinfo {author}
  {\bibfnamefont {T.}~\bibnamefont {Carvalho}}, \bibinfo {author}
  {\bibfnamefont {S.}~\bibnamefont {Roorda}}, \bibinfo {author} {\bibfnamefont
  {M.}~\bibnamefont {Hejna}}, \bibinfo {author} {\bibfnamefont
  {S.}~\bibnamefont {Torquato}}, \ and\ \bibinfo {author} {\bibfnamefont
  {P.~J.}\ \bibnamefont {Steinhardt}},\ }\bibfield  {title} {\enquote {\bibinfo
  {title} {{Hyperuniformity in amorphous silicon based on the measurement of
  the infinite-wavelength limit of the structure factor}},}\ }\href {\doibase
  10.1073/pnas.1220106110} {\bibfield  {journal} {\bibinfo  {journal} {Proc
  Natl Acad Sci USA}\ }\textbf {\bibinfo {volume} {110}},\ \bibinfo {pages}
  {13250--13254} (\bibinfo {year} {2013})}\BibitemShut {NoStop}%
\bibitem [{\citenamefont {Jiao}\ \emph {et~al.}(2014)\citenamefont {Jiao},
  \citenamefont {Lau}, \citenamefont {Hatzikirou}, \citenamefont
  {Meyer-Hermann}, \citenamefont {Corbo},\ and\ \citenamefont
  {Torquato}}]{Jiao2014}%
  \BibitemOpen
  \bibfield  {author} {\bibinfo {author} {\bibfnamefont {Y.}~\bibnamefont
  {Jiao}}, \bibinfo {author} {\bibfnamefont {T.}~\bibnamefont {Lau}}, \bibinfo
  {author} {\bibfnamefont {H.}~\bibnamefont {Hatzikirou}}, \bibinfo {author}
  {\bibfnamefont {M.}~\bibnamefont {Meyer-Hermann}}, \bibinfo {author}
  {\bibfnamefont {J.~C.}\ \bibnamefont {Corbo}}, \ and\ \bibinfo {author}
  {\bibfnamefont {S.}~\bibnamefont {Torquato}},\ }\bibfield  {title} {\enquote
  {\bibinfo {title} {{Avian photoreceptor patterns represent a disordered
  hyperuniform solution to a multiscale packing problem}},}\ }\href {\doibase
  10.1103/PhysRevE.89.022721} {\bibfield  {journal} {\bibinfo  {journal} {Phys
  Rev E}\ }\textbf {\bibinfo {volume} {89}},\ \bibinfo {pages} {022721}
  (\bibinfo {year} {2014})}\BibitemShut {NoStop}%
\bibitem [{\citenamefont {Dreyfus}\ \emph {et~al.}(2015)\citenamefont
  {Dreyfus}, \citenamefont {Xu}, \citenamefont {Still}, \citenamefont {Hough},
  \citenamefont {Yodh},\ and\ \citenamefont {Torquato}}]{Dreyfus2015}%
  \BibitemOpen
  \bibfield  {author} {\bibinfo {author} {\bibfnamefont {R.}~\bibnamefont
  {Dreyfus}}, \bibinfo {author} {\bibfnamefont {Y.}~\bibnamefont {Xu}},
  \bibinfo {author} {\bibfnamefont {T.}~\bibnamefont {Still}}, \bibinfo
  {author} {\bibfnamefont {L.~A.}\ \bibnamefont {Hough}}, \bibinfo {author}
  {\bibfnamefont {A.~G.}\ \bibnamefont {Yodh}}, \ and\ \bibinfo {author}
  {\bibfnamefont {S.}~\bibnamefont {Torquato}},\ }\bibfield  {title} {\enquote
  {\bibinfo {title} {{Diagnosing hyperuniformity in two-dimensional,
  disordered, jammed packings of soft spheres}},}\ }\href {\doibase
  10.1103/PhysRevE.91.012302} {\bibfield  {journal} {\bibinfo  {journal} {Phys
  Rev E}\ }\textbf {\bibinfo {volume} {91}},\ \bibinfo {pages} {12302}
  (\bibinfo {year} {2015})}\BibitemShut {NoStop}%
\bibitem [{\citenamefont {Skoge}\ \emph {et~al.}(2006)\citenamefont {Skoge},
  \citenamefont {Donev}, \citenamefont {Stillinger},\ and\ \citenamefont
  {Torquato}}]{skoge2006}%
  \BibitemOpen
  \bibfield  {author} {\bibinfo {author} {\bibfnamefont {M.}~\bibnamefont
  {Skoge}}, \bibinfo {author} {\bibfnamefont {A.}~\bibnamefont {Donev}},
  \bibinfo {author} {\bibfnamefont {F.~H.}\ \bibnamefont {Stillinger}}, \ and\
  \bibinfo {author} {\bibfnamefont {S.}~\bibnamefont {Torquato}},\ }\bibfield
  {title} {\enquote {\bibinfo {title} {{Packing hyperspheres in
  high-dimensional {E}uclidean spaces}},}\ }\href {\doibase
  10.1103/PhysRevE.74.041127} {\bibfield  {journal} {\bibinfo  {journal} {Phys
  Rev E}\ }\textbf {\bibinfo {volume} {74}},\ \bibinfo {pages} {41127}
  (\bibinfo {year} {2006})}\BibitemShut {NoStop}%
\bibitem [{\citenamefont {Savitzky}\ and\ \citenamefont
  {Golay}(1964)}]{savitzky64}%
  \BibitemOpen
  \bibfield  {author} {\bibinfo {author} {\bibfnamefont {A.}~\bibnamefont
  {Savitzky}}\ and\ \bibinfo {author} {\bibfnamefont {M.~J.~E.}\ \bibnamefont
  {Golay}},\ }\bibfield  {title} {\enquote {\bibinfo {title} {{Smoothing and
  Differentiation of Data by Simplified Least Squares Procedures}},}\ }\href
  {\doibase 10.1021/ac60214a047} {\bibfield  {journal} {\bibinfo  {journal}
  {Anal Chem}\ }\textbf {\bibinfo {volume} {36}},\ \bibinfo {pages}
  {1627--1639} (\bibinfo {year} {1964})}\BibitemShut {NoStop}%
\bibitem [{\citenamefont {Ruh}\ \emph {et~al.}(2018)\citenamefont {Ruh},
  \citenamefont {Scherer},\ and\ \citenamefont {Kiselev}}]{ruh2018}%
  \BibitemOpen
  \bibfield  {author} {\bibinfo {author} {\bibfnamefont {A.}~\bibnamefont
  {Ruh}}, \bibinfo {author} {\bibfnamefont {H.}~\bibnamefont {Scherer}}, \ and\
  \bibinfo {author} {\bibfnamefont {V.~G.}\ \bibnamefont {Kiselev}},\
  }\bibfield  {title} {\enquote {\bibinfo {title} {{The Larmor Frequency Shift
  in Magnetically Heterogeneous Media Depends on Their Mesoscopic
  Structure}},}\ }\href {\doibase 10.1002/mrm.26753} {\bibfield  {journal}
  {\bibinfo  {journal} {Magn Reson Med}\ }\textbf {\bibinfo {volume} {79}},\
  \bibinfo {pages} {1101--1110} (\bibinfo {year} {2018})}\BibitemShut {NoStop}%
\end{thebibliography}%

\subsection*{Acknowledgements}
AR, PE and VGK were supported by the German Research Foundation (DFG), grant KI\,1089/6-1. 
DSN was supported in part by the Center of Advanced Imaging Innovation and Research (CAI2R, www.cai2r.net), 
a NIH/NIBIB Biomedical Technology Research Center: P41 EB017183, and by NIH/NINDS grants R01 NS088040 and R01 NS039135. %The most crucial support of his work was offered by numerous breweries in Freiburg, Germany. 

%\subsection*{Author contrubutions}
%The study was designed by DSN and VGK who also developed analytical theory. Numerical simulations were performed by AR and VGK, experiments were performed by AR, PE and HS, and the manuscript was written by AR, DSN and VGK.

%\subsection*{Additional information}
%Authors declare no competing financial interests. 

%%%%%%%%%%%%%%%%%%%%%%%%%%%%%%%%%%%%%%%%%%%%%%%%%%%%%%%%%%%%%%%%%%%%%%%%%%%%
%\newpage

\small
\section*{Methods}

{\bf Synthetic media.}
In \fig{fcorrdR2}, we consider five representative media: four types of differently arranged identical spheres with radius $\rho$, and one with randomly placed long (prolate) ellipsoids (here $\rho$ is the radius of the short axes, the long semi-axis is $40\rho$), which were generated as follows. A simple cubic lattice of spheres with a volume fraction of $\zeta = 34\%$ represents perfect order. To create the shuffled lattice, the spheres of the regular lattice were randomly displaced from their original positions by nine discrete values within the distance $\pm 0.2335\rho$ in each direction rejecting steps that caused overlap with a neighbor. The MRJ packing was generated using an event-driven molecular dynamics simulation \cite{skoge2006} using the code dowloaded from the authors' website. The resulted medium had the volume fraction of spheres $\zeta = 65\%$. For the short-range disorder, the spheres were randomly added rejecting steps leading to the overlap with already existing spheres ($\zeta = 34\%$). The same algorithm was used for the random arrangement of long ellipsoids whereas the non-overlap condition was released resulting the ellipsoids with the summed volume of $15\%$ of the simulation volume forming a structure with the overall volume fraction $\zeta = 14\%$. To alleviate finite-size effects in the diverging power spectrum of the long ellipsoids at small $k$ (\fig{fcorrdR2}), we further averaged the MC runs over ten different disorder realizations. All media were sampled on a $1024^3$ cubic grid for numerical computation of $\Omega(\r)$ and successive MC simulations.

The Larmor frequency shift, $\Omega(\r)$, was calculated as the convolution with the elementary dipole field. To characterize the scale (the dephasing strength) of field variations, we use the dephasing introduced by a single object: $\delta \Omega = \frac43 \pi \chi \gamma B_0 $ for spheres and $\delta \Omega = 2\pi \chi \gamma B_0$ for the long ellipsoids, where $ \chi$ is the susceptibility difference with the background, and $\gamma B_0$ the Larmor frequency in the external field. The disorder correlation length $l_c$ was defined starting with $k_c=\pi/\rho$, which is close to the pronounced peak of $\overline \Gamma_2(k)$ for the considered sphere packings, correspondingly $l_c = 1/k_c$ and the correlation time $t_c = 1 \,/\, (D \, k_c^2)$. The dimensionless parameter $\alpha = \delta \Omega \, t_c$ instantiates the typical spin phase, $\langle \varphi_1^2\rangle$, acquired when moving over the disorder correlation length; the diffusion-narrowing takes place when $\alpha\ll 1$. 

{\bf Numerical calculations} to obtain $\Sigma(\omega)$ and $\d R_2(t)/\d t$ to the leading order were performed by integrating the Larmor frequency power spectra $\Gamma_2(\k) = \Omega_{-\k} \Omega_\k/V$, cf. \eq{Gamma_k}, for the five synthetic media according to \eqsand{sigCorrection}{dR2pert}, respectively, while neglecting $\Rinf$ on the right-hand side.

{\bf Monte Carlo simulations} of the mesoscopic relaxation for freely diffusing spins were performed with $N_s = 10^8$ spins randomly hopping on the sample grid of the above described media imitating the dephasing strength $\alpha = 0.2$ (spheres) and $\alpha = 0.05$ (long ellipsoids). The increment of the random walker's spin phase was calculated using the mean of $\Omega(\r)$ before and after each hop. The mesoscopic NMR signal at each time moment $t$ was calculated as the mean of all accumulated phase factors $s(t) = \langle e^{-i \varphi(t)} \rangle$. The second derivative of $\ln s(t)$ was calculated using third-order polynomial fitting, based on Savitzky-Golay filtering,\cite{savitzky64} with a linearly increasing filter width of $0.6 \, t$.

{\bf Microbead samples.} 
Polystyrene microbeads (Dynoseed TS10; Microbeads AS, Skedsmokorset, Norway) were suspended in an aqueous solution of sodium chloride doped with Holmium(III) chloride hexahydrate ($\mathrm{HoCl_3 \cdot 6H_2O}$) in various concentrations to adjust the solution density and magnetic susceptibility, respectively.\cite{ruh2018} Suspensions with 30\% volume fraction of microbeads were prepared using a particle-density matched sodium chloride solution ($c_\mathrm{NaCl}=1.28 \units{mol/L}$ for $T = 309 \units{K}$) to avoid sedimentation. MRJ samples were prepared by particle sedimentation in $c_\mathrm{NaCl}=0.33 \units{mol/L}$ solution and careful removal of particle-free fluid from the top. All samples were prepared in standard 5-mm NMR tubes.

{\bf NMR measurements} were performed on a DPX $200\units{MHz}$ spectrometer (Bruker, Ettlingen, Germany) using a standard zg30 sequence (flip angle $=30^\circ$, acquisition time $=4\units{s}$, 16 averages, relaxation delay $=3 \units{s}$, no spinning) at $T = 309 \units{K}$. The shim fields were adjusted on a pure $\mathrm{D_2O}$ sample and then kept for all samples within a measurement series. To obtain $\d R_2/\d t$, the measured FID signals were processed with the same fitting algorithm\cite{savitzky64} as applied for MC simulations using a filter width of $0.7 \, t$.

%%%%%%%%%%%%%%%%%%%%%%%%%%%%%%%%%%%%%%%%%%%%%%%%%%%%%%%%%%%%%%%%%%%%%%%%%%%%
\clearpage
\normalsize

\counterwithin{figure}{section}
\counterwithin{equation}{section}
\renewcommand{\theequation}{S\arabic{equation}}
\renewcommand{\thefigure}{S\arabic{figure}}
\setcounter{equation}{0}
\setcounter{figure}{0}

\section*{Supplemental Information}
\begin{figure}[h!] 
\includegraphics[width=\textwidth]{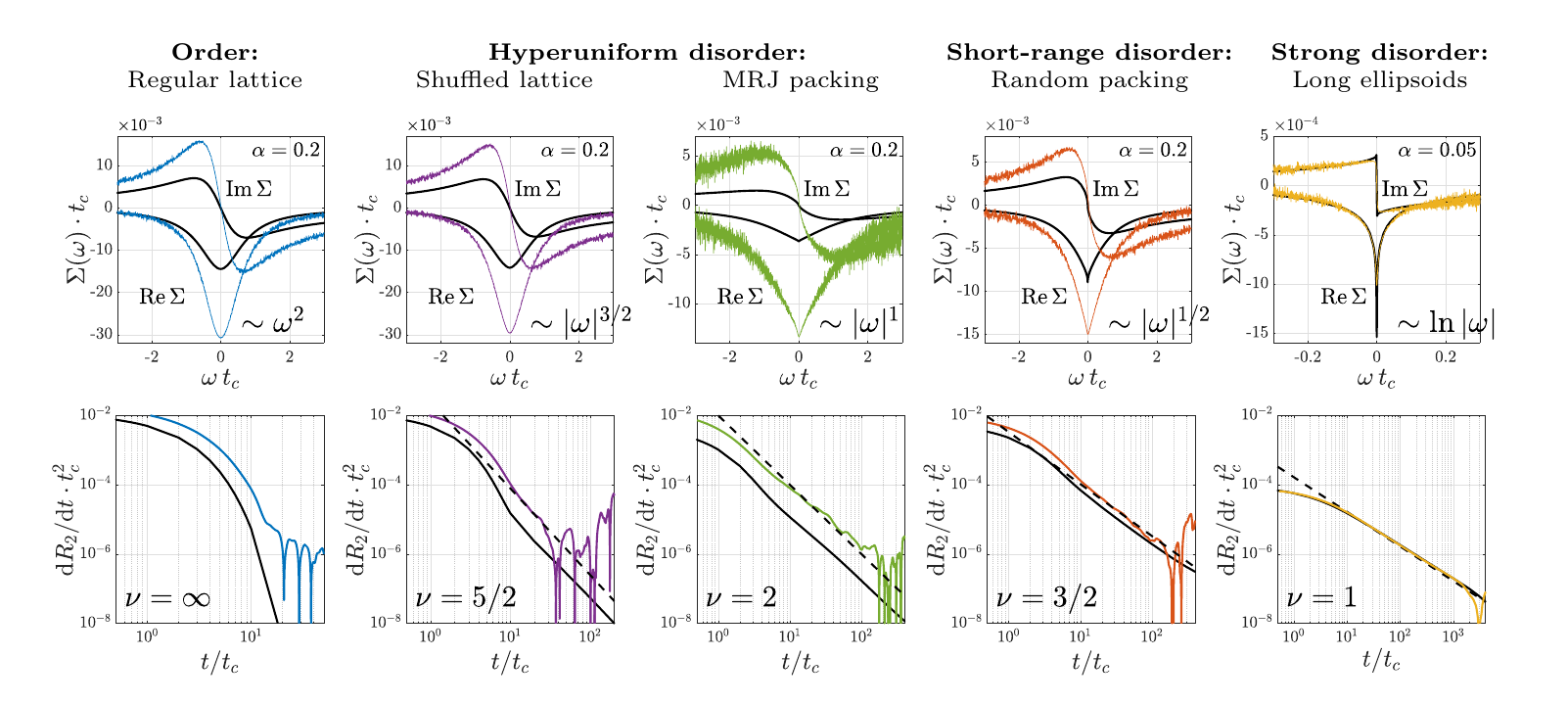}
\caption{
Results for hindered diffusion in the five three-dimensional synthetic media from \fig{fcorrdR2} representing different disorder classes. The second row shows the derivative of the time-dependent relaxation rate, $\d R_2 / \d t$. MC results for hindered diffusion (colored curves) in general differ from the leading-order calculation (black curves, the same as in \fig{fcorrdR2}), which assumes permeable objects, but follow the same characteristic exponents (dashed lines). The same observation holds for the equivalent depiction in frequency domain (first row). While the self-energy curves also differ quantitatively from the leading-order calculation, the qualitative behavior embodied by the power-law exponent at $\w=0$ is similar. For strong disorder the difference between hindered and free diffusion is actually negligible, which follows from the low volume fraction of $\zeta=14\%$ for the long ellipsoids.
\label{imperfigs} 
}
\end{figure}

\begin{figure}[h!]
\includegraphics[height=5cm]{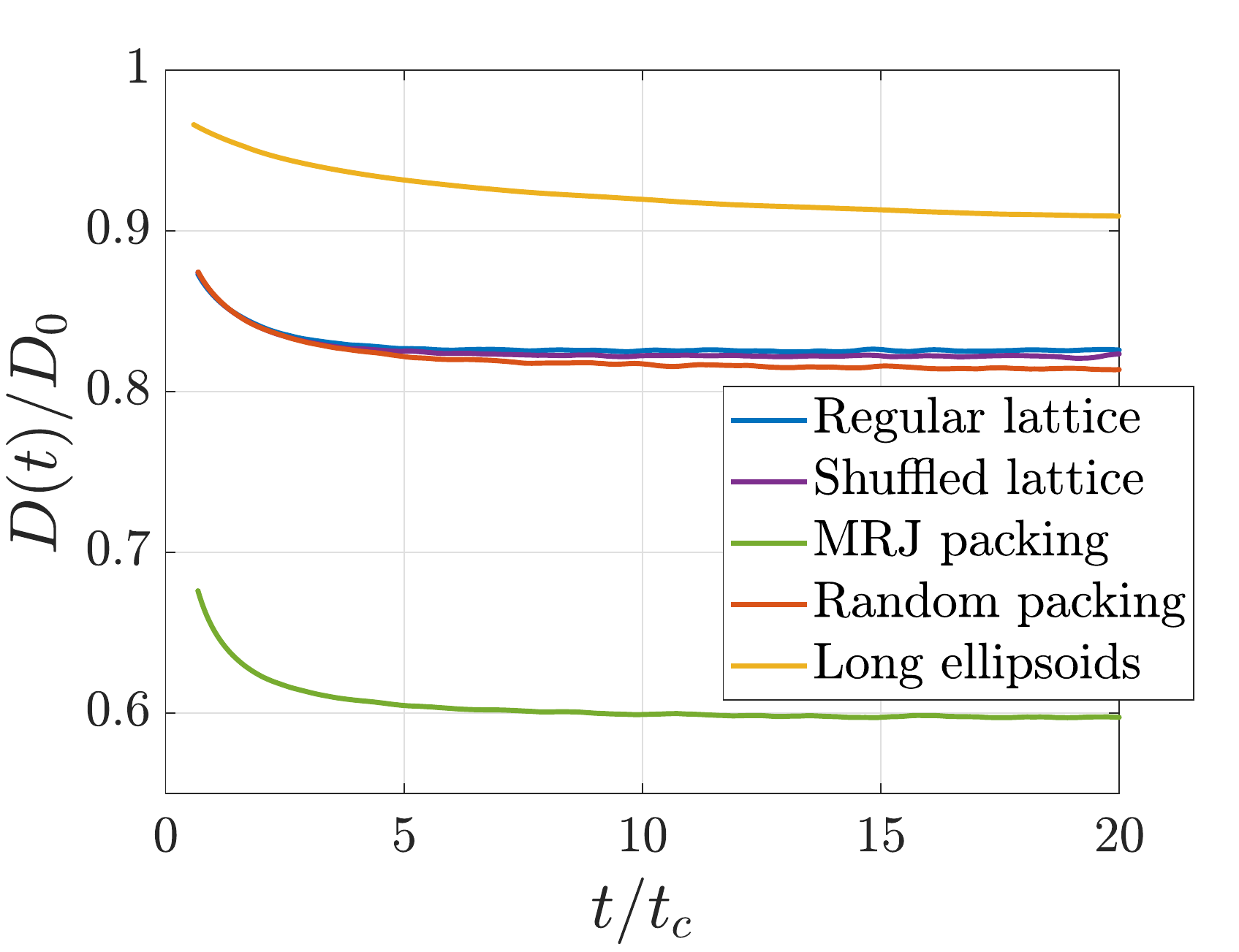}
\caption{Instantaneous diffusion coefficient\cite{novikov2010} $D(t) = \frac16 \partial_t \langle x^2(t)\rangle$ for the hindered diffusion within the investigated media, normalized to the free diffusion coefficient $D_0$. The temporal derivative of the mean square displacement $\langle x^2(t)\rangle$ of the MC random walkers was computed using the Savitzky-Golay filter\cite{savitzky64} based on the second-order polynomial, with a width of about $t_c$.
\label{diffcoef}
}
\end{figure}

%\subsection{Results for hindered diffusion}
Monte Carlo simulations for hindered diffusion were performed within the same media and with the same parameters as in \fig{fcorrdR2}. Impermeable spheres and ellipsoids were simulated by discarding Monte Carlo steps that lead inside the objects in which case the random walkers did not move during the given time step. The results shown in \fig{imperfigs} support the universality of the dynamical exponent, while the coefficients in front of $t^{-\nu}$ and $|\omega|^{\nu-1}$ are non-universal. Note that the renormalization of the diffusion constant with its long-time asymptote (\fig{diffcoef}) is not sufficient to reproduce the non-universal coefficients (data not shown), which are strongly modified for the media with higher volume fraction $\zeta$.

\onecolumngrid
%%%%%%%%%%%%%%%%%%%%%%%
\vspace{4em}

%%%%%%%%%%%%%%%%%%%%%%%

\end{document}